# Spatial Variations of Jovian Tropospheric Ammonia via Ground-Based Imaging

S. M. Hill[1], P. G. J. Irwin[2], C. Alexander[2], and J. H. Rogers[3]

[1]Independent Researcher, Denver, CO 80224, USA.

[2]Atmospheric, Oceanic and Planetary Physics, Department of Physics, University of Oxford, Parks Rd, Oxford, OX1 3PU, UK.

[3]British Astronomical Association, London, UK.

**Key Points:**

- Ground-based, narrow-band imaging with a small telescope reveals spatially resolved ammonia abundance in Jupiter's upper troposphere.

- Values from disk-averages, meridional profiles, and two-dimensional maps compare favorably with large instrument results.

- Small observatory observations can offer frequent observations to provide context and continuity for professional research programs.

Corresponding author: Steven M. Hill (smhill001@gmail.com)

**Abstract**

Current understanding of the ammonia distribution in Jupiter's atmosphere is provided by observations from major ground-based facilities and spacecraft, and analysed with sophisticated retrieval models that recover high fidelity information, but are limited in spatial and temporal coverage. Here we show that the ammonia abundance in Jupiter's upper troposphere, which tracks the overturning atmospheric circulation, can be simply, but reliably determined from continuum-divided ammonia and methane absorption-band images made with a moderate-sized Schmidt-Cassegrain telescope (SCT). In 2020-21, Jupiter was imaged in the 647-nm ammonia absorption band and adjacent continuum bands with a 0.28-m SCT, demonstrating that the spatially-resolved ammonia optical depth could be determined with such a telescope. In 2022-23, a 619 nm methane-band filter was added to provide a constant reference against which to correct the ammonia abundances (column-averaged mole fraction) for cloud opacity variations. These 0.28-m SCT results are compared with observations from: a) the MUSE instrument on ESO's Very Large Telescope (VLT) b) TEXES mid-infrared spectrometer used on the NASA's InfraRed Telescope Facility (IRTF); and c) the Gemini telescopes, and are shown to provide reliable maps of ammonia abundance. Meridional and longitudinal features are examined, including the Equatorial Zone (EZ) ammonia enhancement, the North Equatorial Belt (NEB) depletion, depletion above the Great Red Spot (GRS), and longitudinal enhancements in the northern EZ. This work demonstrates meaningful ammonia monitoring can be achieved with small telescopes that can complement spacecraft and major ground-based facility observations.

**Plain Language Summary**

On Jupiter, ammonia – in addition to water – forms clouds, precipitates, and evaporates. The main clouds we see are commonly thought to be made of ammonia ice, with 'coloring agents' that provide sometimes vivid hues. Where ammonia is and is not provides a powerful tracer of weather processes on Jupiter, making it important for understanding the planet and others like it. Most measurements of ammonia are made using large astronomical facilities or spacecraft. However, because of competition for observing time at large telescope facilities and the limited views of spacecraft, professional observations may only occur a few times a year or only cover part of Jupiter. We show here that useful measurements of ammonia can be made with small telescopes and equipment available to many amateur astronomers. By comparing images that show absorption due to ammonia and methane gases, we can determine the variation in the amount of ammonia in and above Jupiter's clouds. We have compared our small telescope measurements with those from large professional telescopes and found similar results on the distribution of ammonia. Our observation method means that non-professional observers can fill observational gaps and provide context for professional observing campaigns.

**1 Introduction**

Ammonia vapor serves as an important tracer of Jovian tropospheric meteorology. With the advent of new observations over the past decade, spanning visible to radio wavelengths, there has been a renaissance in studies of Jovian meteorology, including by ground-based observations [e.g., *Bjoraker et al.,* 2022, *Braude et al.*, 2017; *Braude et al.*, 2020; *Dahl et al.*, 2021; *de Pater et al.,* 2023; *Fletcher et al.*, 2016; *Fletcher et al.*, 2020; *Fletcher et al.*, 2017; *Fletcher et al.*, 2021, *Wong et al.,* 2023]. Sophisticated radiative transfer and atmospheric retrieval codes, e.g., NEMESIS (*Irwin et al*., 2008), can obtain aerosol and trace gas properties to depths of 100 bars using JUNO Microwave Radiometer (MWR) data, including the ammonia distribution [Bolton *et*

*al.*, 2017, *Li et al.*, 2017, *Guillot et al.*, 2020]. Most of these studies require significant financial, human, facility, and computational resources, which limits observational coverage. For ground-based facilities there are additional factors that limit temporal coverage and revisit time such as competition for telescope time and adverse weather or seeing conditions. Planetary spacecraft face factors that limit contemporaneous spatial coverage such as their proximity to the planet and finite instrument field-of-views.

The current investigation was prompted by the potential to fill observational gaps in professional investigations of the upper tropospheric ammonia distribution [*Hill and Rogers*, 2022]. By examining the ratio of ammonia to methane absorption, the abundance of ammonia can be estimated since the abundance of methane is fixed and well known (mole fraction = $1.81 \times 10^{-3}$) [*Niemann et al.*, 1998]. We focus on a relatively narrow portion of the red spectrum to minimize differences in scattering paths and properties in the Jovian atmosphere. Specific guidelines for applying this technique to determine relative gas abundances have been previously described in the literature [*Combes and Encrenaz*, 1979]. Spectroscopy has long been used to measure abundances and their spatial variations [*Lutz and Owen*, 1980; *Woodman et al.*, 1977]. Woodman et al. [1977] state, "We may have developed a technique for studying local weather on Jupiter, rather than a means for determining the planet's chemistry." Here, we extend this technique to two dimensions through narrow-band imaging specifically to study Jovian weather.

In Section 2, this paper describes the observations and equipment used. In Section 3, it presents the methodology to obtain absorption and abundance measurements for ammonia. In the Sections 4-6, results are described for the integrated disk, meridional profiles, and 2-D maps respectively. In each case, the results of this work are compared to other previously published works. Section 7 provides a summary and conclusions.

**2 Observations**

2.1 0.28-m SCT Observations

Observations were carried out from a private observatory in Denver, Colorado, USA (Table 1). All equipment was commercially obtained and included a Celestron 0.28-m Schmidt-Cassegrain telescope (henceforth 'SCT'), a ZWO ASI120MM CMOS camera, and an iOptron CEM70 mount. The imaging configuration used obtains image sequences at a high cadence, 100 ms or faster. The observations were made at a long focal length (plate scale of 0.25 arcsec-pix$^{-1}$) to detect discrete ammonia abundance features.

**Table 1.** SCT observations carried out from the 2020 apparition through the 2023 apparition. In 2020 and 2021, only opacity is observed for ammonia. In 2022 and 2023, with the addition of the 619 nm and 632 filters, methane opacity is also observed, permitting the determination of ammonia abundance.

| Apparition | Observed Date Range | Num. Obs. | 619 nm | 632 nm | 647 nm | 656 nm | NH$_3$ Parameters |
|---|---|---|---|---|---|---|---|
| 2020 | Jul 20 – Jul 31 | 4 | - | - | X | X | Opacity |
| 2021 | Jun 22 – Dec 3 | 19 | - | X* | X | X | Opacity |
| 2022 | Aug 10 – Jan 13 | 19 | X | X | X | X | Opacity and abundance |
| 2023 | Aug 15 – Mar 1 | 80 | X | X | X | X | Opacity and abundance |

* The first observation of the 2021 season only included the 656 nm filter for continuum measurement.

A typical observing session lasts between 20 and 30 minutes with the objective of minimizing the offset of central observation times in each band (Table 2). This is done by taking nested pairs of image sequences in each 'science' band (619, 632, 647, and 656 nm) which are centered on the same overall observation time. Typically, the best 25% of images are selected, aligned, and stacked using RegiStax software in a process described 'lucky imaging.' Selection is based on sharpness criteria and stacking this subset of sharper images results in better resolution on the final co-added image from the sequence than co-adding the entire sequence. This technique is widely used by amateur astronomers for planetary imaging. With an image sequence duration of 2 minutes, this results in integration times of about 30 seconds per filter. Before analysis, each image pair needs to be rotated (with respect to the central meridian longitude) to the central time and stacked (averaged). This is done using WINJupos software (https://jupos.org) with the default limb darkening coefficient of 1.0 (simple Lambertian scattering). The RGB context image (usually IR-GB for better resolution and throughput in the red channel) is taken after the science data.

Table 2 shows an idealized observing sequence of 2-minute sequences in the methane, ammonia, and continuum filters, which are spatially binned over 2x2 pixel (0.5×0.5 arcsec) boxes, followed by the 1-minute RGB sequences, which are unbinned. In reality there is overhead time (total <5 minutes) for filter changes and focus checks. In total, This approach means that minimal adjustment is needed between the stacked, derotated science bands, though in the creation of each stacked image up to ~4.5 deg of planetary rotation needs to be corrected for (performed, as already noted, using the derotation function of WinJUPOS). Because the context image is not nested, it has an additional rotational offset, but that is deemed less important since it is not used for quantitative analysis.

**Table 2.** Idealized set of sequences for a single observation. The 'nesting' strategy minimizes planetary derotation to a common Central Meridian Longitude. Total number of image frames is 23,600, of which approximately 6,900 are selected for co-addition within their spectral bands.

| Purpose | Center Wavelength (nm) | Sequence Duration (s) | Binning | Typical Exposure (ms) | Number of Images | Center Time (s) | CM II (deg) |
|---|---|---|---|---|---|---|---|
| Continuum | 656 | 120 | 2x2 | 100 | 1200 | -420 | -4.2 |
| Continuum | 632 | 120 | 2x2 | 50 | 2400 | -300 | -3.0 |
| CH4 | 619 | 120 | 2x2 | 50 | 2400 | -180 | -1.8 |
| NH3 | 647 | 120 | 2x2 | 50 | 2400 | -60 | -0.6 |
| NH3 | 647 | 120 | 2x2 | 50 | 2400 | 60 | 0.6 |
| CH4 | 619 | 120 | 2x2 | 50 | 2400 | 180 | 1.8 |
| Continuum | 632 | 120 | 2x2 | 50 | 2400 | 300 | 3.0 |
| Continuum | 656 | 120 | 2x2 | 100 | 1200 | 420 | 4.2 |
| (IR)GB Context | >685 | 120 | 1x1 | 50 | 2400 | 510 | 5.1 |
| (IR)GB Context | 550 | 60 | 1x1 | 30 | 2000 | 570 | 5.7 |
| (IR)GB Context | 450 | 120 | 1x1 | 50 | 2400 | 630 | 6.3 |

### 2.2 VLT/MUSE Observations

The SCT observations are compared with observations made using the MUSE instrument mounted on the VLT Unit 4 telescope [*Bacon et al.*, 2010]. MUSE has previously been used for cloud and ammonia retrievals using NEMESIS [*Braude et al.*, 2020; *Irwin et al.*, 2018, 2019a, 2019b].

The VLT has an aperture of 8 m and the MUSE instrument has a seeing-limited plate scale of 0.2 arcsec-pix$^{-1}$ in Wide Field Mode, similar to that of the CMOS lucky images for the 0.28-m SCT observations. However, MUSE simultaneously samples 3722 spectral bands from 470 to 935 nm with a spectral resolution of ~0.26 nm. Two MUSE observations are considered: 2022 Jul. 30, 0730UT, roughly midway between PJ43 and PJ44 and 2022 Sep. 19, 0352UT, affiliated with the Juno PJ45 (September 29) campaign. This work highlights the second observation because it is only an hour before a 0.28 m SCT observation at 0453UT. The MUSE radiance data cubes were calibrated and corrected for telluric absorption (Braude, 2019; Braude *et al.*, 2020). Additionally, the radiance was converted to reflectivity using the the Chance and Kurcuz (2010) reference solar spectrum.

The MUSE radiance data cubes were averaged over Jupiter's disk at each wavelength, then divided by the incident solar flux spectrum (Chance and Kurcuz, 2010) to obtain reflectivity in units of *I/F*. The resulting spectra are plotted in Figure 1a and compare favorably with the Karkoschka (1994) observation. Differences between the three spectra may be due to changes in Jupiter's atmosphere between the observations. Figure 1b shows the SCT filter normalized transmission profiles and Fig. 1c shows the absorption coefficients for methane and ammonia, which will be discussed further in Section 3.

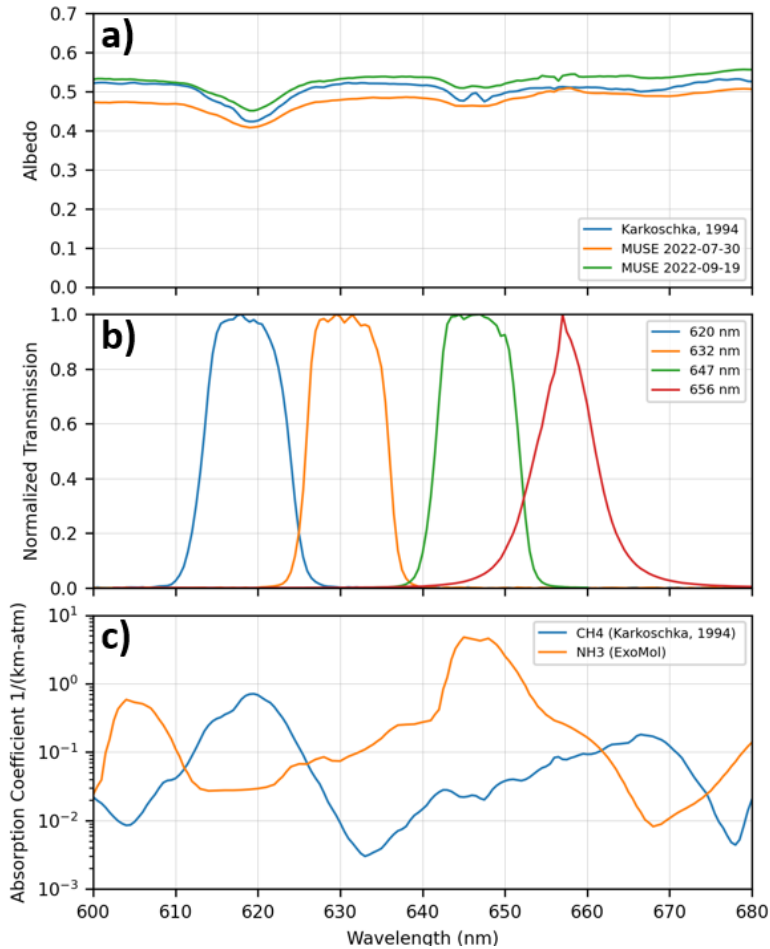

**Figure 1.** Disk-averaged reflectivity, filter transmission profiles, and molecular absorption coefficients are shown. a) Jovian reflectivity measured by Karkoschka (1994) and from two VLT/MUSE observations in 2022. b) Filter transmission, normalized to the peak of each band, for the filters used in the SCT observations and applied to the VLT/MUSE data cubes. c) Molecular absorption coefficients for methane and ammonia (see text for details).

In order to compare SCT images to MUSE observations, four images for each MUSE observing date were created by convolving the data cubes with the SCT filter profiles. To provide MUSE context images, an RGB image was created by stacking spectral images from 475-480 nm ("blue"), 530-580 nm (green), and 630-680 nm (red).

## 3 Methodology

The starting point for the analysis is to obtain 'images' of Jupiter that represent the two-way absorption in the 619 nm methane band and in the 647 nm ammonia band. Specific guidelines for applying this technique to determine relative gas abundances have been described in the literature by Combes and Encrenaz [1979], namely:
1) The two lines or bands have to be chosen in the same spectral range.
2) The observed flux is roughly equal for both absorbers.
3) The ratio that we want to determine is constant with height.

4) The two absorption coefficients have the same dependence with height, which means the same dependence on temperature and pressure.

Thus the first two criteria are satisfied for this use of the 619 and 647 nm absorption bands, as they have similar depth (two-way transmission of 50% at ~10 bars in a scattering-free atmosphere), and penetrate to similar atmospheric levels. However, Combes and Encrenaz [1979] noted particular issues in applying the technique for measuring the $NH_3/CH_4$ ratio, because ammonia's condensation results in very different vertical profiles for the two gases above the ammonia condensation level at ~0.7 bar. In this work, $CH_4$ is already known and we are specifically investigating the horizontal spatial variation of ammonia, not vertical; thus, criterion 3) provides a caution this interpretation due to possible confusion between vertical and horizontal variations. For the final criterion, the absorption coefficients used do not vary hugely over the pressure and temperatures estimated in the upper troposphere of Jupiter ($< 5 - 10\%$) and so are here approximated to be constant.

Here we create an effective continuum image at 647 nm for the ammonia band by linear interpolation on a pixel-by-pixel basis between the 632 nm and 656 nm continuum images. Wong et al. [2023] independently support the choice of 623 nm as a continuum wavelength in their Fig. 13. The use of the 656 nm image helps offset minor contamination of the 647 nm ammonia band by the methane band at 668 nm. The 632 nm image is used on its own for the 619 nm methane band continuum. Note that for the SCT data we use the directly observed images in each filter, whereas for MUSE, we are using images created by convolving the filter profiles (Figure 1b) with the data cubes as described in Section 2. We then divide the absorption band images by the continuum images to obtain the mean transmission, $T$, over the methane and ammonia absorption bands:

$$T_{CH4} = \frac{R_{619}}{R_{632}}, T_{NH3} = \frac{R_{647}}{((1-x)R_{632} + xR_{656})} \quad (1)$$

where $R_{619}$ is the observed mean reflectivity in the 619-nm filter, etc., and the denominator for $T_{NH3}$ is an estimate of the continuum at the centre of the ammonia absorption band, with $x = (647 - 632)/(656 - 632)$. The opacities can then be calculated as $\tau_{CH4} = -\ln(T_{CH4})$ and $\tau_{NH3} = -\ln(T_{NH3})$.

The expected absorption of a gas can be calculated as:

$$\tau_{gas} = \rho_{gas} s k_{eff} \quad (2)$$

where $\rho$ is the absorbing gas density (amagat), $s$ is the two-way scattering path length (absorption column length, km), and $k_{eff}$ ((km-amagat)$^{-1}$) is the effective absorption coefficient of the gas, averaged over the relevant filter function. Note that an amagat is the number density of an ideal gas at 1 atm pressure (i.e., 101.325 kPa) and 0°C (i.e., 273.15 K), equal to 2.68678×10$^{25}$ molecule/m$^3$, or 44.165 mol/m$^3$. The mole fraction is $f_{gas} = n_{gas} / n$, where $n$ is the total number density of all atmospheric constituents.

Table 3 shows the computed $k_{eff}$ values for methane and ammonia calculated for the different filters considered using the filter profiles from Figure 1b and the absorption coefficients from Figure 1c. For methane we assume the absorption coefficient spectrum of Karkoschka [1994], while for ammonia we use a spectrum calculated from a k-table (Lacis and Oinas, 1991). fitted by Irwin et al. [2019a] to the ExoMOL data of Coles et al. [2019], computed for typical conditions of 1 atm and 165 K. The implications of the small, but nonzero, relative values of the continuum filters compared to the in-band filters is discussed in Section 4.

**Table 3.** Effective Absorption Coefficients, $k_{eff}$ (km-amagat)$^{-1}$, for the SCT filters profiles. Karkoschka [1994] cross sections were used for methane and Coles et al. [2019] line data, reduced to a k-table by Irwin et al. [2019] were used for ammonia, calculated at a temperature of 165 K and pressure of 1 atm.

|  | $CH_4$ | $NH_3$ |
|---|---|---|
| 619 $CH_4$ | 0.427 | 0.033 |
| 632 Continuum | 0.019 | 0.103 |
| 647 $NH_3$ | 0.028 | 2.955 |
| 656 Continuum | 0.079 | 0.479 |

Images at 619 nm and 647 nm, divided by their associated continuum image, reveal the strength of absorption in each gas, and therefore provide estimates of the integrated column abundance above the cloud-tops. Figure 2 shows this process for methane (a-c) and ammonia (d-f), progressing from the in-band radiance images, then to the computed transmission images, and finally to the opacity images. The effective cloud top pressures, a visual image for orientation, and the ammonia abundance are shown in Figure 2g-i.

The absorption alone cannot convey the ammonia concentration as we also need to correct out variations in the clouds. The cloud-top pressure can be estimated independently from images in methane absorption band(s), as methane is uniformly distributed in Jupiter's atmosphere. The well-known strong absorption band at 889 nm is often used to give a qualitative estimate of cloud-top pressures. However, the 889 nm band violates the Combes and Encrenaz [1979] criterion 1) by not being spectrally close to the 647 nm ammonia band and criterion 3) by being much more sensitive to lower pressures. Together, these violations result in the 889 nm band being strongly influenced by high-level hazes as well albedo differences from the red portion of the spectrum. For these reasons we have chosen the 619 nm weak methane band to estimate the cloud top pressure for ammonia abundance analysis.

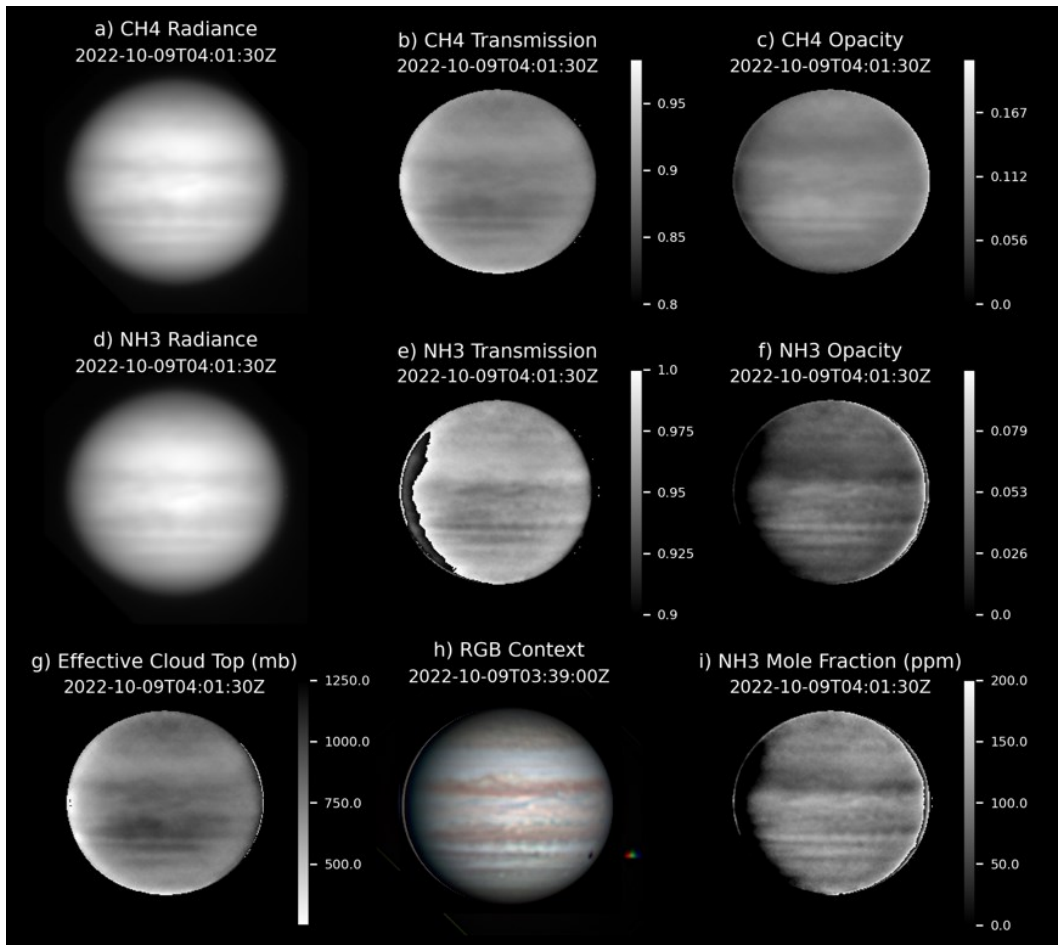

**Figure 2.** Example process of determining effective cloud top pressure and ammonia abundance from an SCT observation 2022-10-09 04:01:30: a) radiance in the 619 nm methane band, b) transmission in the methane band (i.e., radiance in methane band divided by radiance in nearby continuum band), c) methane band opacity, d) radiance in the 647 nm ammonia band, e) transmission in the ammonia band, f) ammonia band opacity, g) effective cloud top pressure (see text for calculation), h) a visual image for orientation (2022-10-09 03:39:00 rotated to the same central meridian longitude at 04:01:30), i) ammonia abundance calculated from the ratio of opacities according to Eq. 6 – brighter areas represent more ammonia relative to methane. Note that the dark artifact at the west limb of e), f), and i) is due to a region where the calculated transmission slightly exceeded unity. This was due to an accidental timing offset between the continuum and absorption images so the calculated transmissions, opacities, and ammonia mole fraction are not valid there. This observation is used as an example, despite the artifact, because it represents the highest spatial resolution achieved in 2022.

The transmission images in Figure 2b and 2e are calibrated by the setting the average pixel value to the disk-integrated transmission values for methane and ammonia respectively, i.e., setting $\frac{1}{N}\sum_1^N T_i = \bar{T}$. Calibration of the SCT data using the Galilean moons as references was explored, but the statistical variations from night to night were significant. Additionally, a systematic offset remained at the 30% level for ammonia and the 10% level for methane. Since the VLT/MUSE data is radiance-calibrated and the disk-average transmissions should not vary significantly over an apparition, we set the SCT disk-average transmission equal to the MUSE average transmissions (Section 4). Opacities computed from the calibrated transmission images can then be used to determine the calibrated pixel-by-pixel cloud top pressure and ammonia abundance.

Since we decouple spatial variations in the data from disk-integrated calibration, uncalibrated (but linear) lucky imaging data can be normalized and then used. This is essential for the use of citizen scientist images because it places less burden on the observer to perform precise calibration. The Jovian disk-integrated ratio of ammonia to methane should be relatively stable, except perhaps during upheavals or other major meteorological events.

From the methane and ammonia opacities, we can compute the estimated total column abundance:

$$\eta N_{gas} = \frac{\tau_{gas}}{k_{eff}} \tag{3}$$

where $N_{gas}$ is the nadir column abundance of the absorbing gas (km-amagat) the reflected sunlight passes through and $\eta$ is an airmass factor for an optically thin continuum that is equal to 2 for disk-center observations and equal to 4 for disk-averaged observations [e.g., *Chamberlain and Hunten*, 1987]. While it is not necessary to compute $N_{gas}$ explicitly to determine the ammonia mole fraction, $N_{CH4}$ can provide useful information on the effective pressure at the cloud tops.

In a Reflecting Layer Model (RLM), the cloud tops are treated as a discrete reflecting surface with a characteristic, wavelength-dependent reflectivity. The column abundance, $X$ (molecule/m²) of a well-mixed gas above a certain pressure level, $p$, is:

$$X = fnH \tag{4}$$

where $f$ is the mole fraction of the gas, $n$ is the number density (molecule/m³) at pressure level $p$ and $H$ is the atmospheric scale height. Substituting for $n$ and $H$ in Eq. 4 we find:

$$X = f \frac{p}{kT} \frac{RT}{\bar{m}g} = \frac{fpN_A}{\bar{m}g} \tag{5}$$

where we have used the gas constant, $R = kN_A$, where $k$ is the Boltzmann constant and $N_A$ is Avogadro's number, and where $\bar{m}$ is the mean molecular weight of the atmosphere (kg/mol) and $g$ is the local gravitational acceleration (m s⁻²). Eq. 5 can be rearranged for pressure, which we take to be the cloud top pressure $P_{cloud}$:

$$p = P_{cloud} = \frac{X\bar{m}g}{fN_A} = 44615 \left(\frac{N_{CH4}}{f_{CH4}}\right) \bar{m}g \tag{6}$$

where $N_{CH4}$ is the column abundance of methane in units of km-amagat, and 44615.0 is the factor to convert from km-amagat to mol/m². Figure 2g shows the effective cloud top pressure, $P_{cloud}$.

To create an image of relative ammonia absorption, we divide the ammonia opacity image (Figure 2f) by the methane opacity image (Figure 2c) to produce an image that represents the ratio of ammonia to methane absorption. To determine the ammonia abundance, we also consider the known absorption coefficients and methane abundance:

$$\bar{f}_c(NH_3) \cong f_{CH4} \frac{\tau_{NH3}}{\tau_{CH4}} \frac{k_{CH4}}{k_{NH3}} \tag{7}$$

where $\bar{f}_c(NH_3)$ is the column-averaged ammonia mole fraction and $f_{CH4}$ is the known mole fraction for methane, (1.81×10⁻³), which is also known to be well-mixed in this pressure region and thus constant with height. Figure 2i shows $\bar{f}_c(NH_3)$ with brighter regions indicating higher ammonia column abundance. Figure 2f shows a visual image for orientation.

Just as path length cancels when computing the gas opacity ratio, geometric factors (illumination and emission angles) also cancel. Thus, we do not need to explicitly calculate path length or geometric factors to obtain $\bar{f}_c(NH_3)$. Nevertheless, the cancellation is not perfect, and we restrict our analysis to within 45º of the central meridian and equator. (Note that geometric factors still apply to explicit calculations on single gas absorption, e.g., path length or derived effective cloud-top pressure.)

While the methane and ammonia transmission images in Fig. 2 may look similar, they are not identical. Thus, the computed ammonia abundance image (Fig. 2i) suggests spatial variations distinct from the cloud top pressure image. Notable features are the increased ammonia near the northern edge of the Equatorial Zone (seen as brighter shades) and the decreased ammonia in the North Equatorial Belt. These features are consistent with reports in the literature as will be discussed in subsequent sections.

## 4 Disk-Integrated Results

Disk-integrated parameters were calculated for the SCT and VLT/MUSE models and observations. Table Table 4 shows the modeled and observed values for transmission, $\bar{T}_{gas}$, optical depth, $\bar{\tau}_{gas}$, and Equivalent Width, $W_{gas}$, of the methane and ammonia absorption bands. Equivalent Width is the spectral width of a band if it was 100% absorbing:

$$W_{gas} = \int_{\lambda_1}^{\lambda_2} (1 - T_{gas}) d\lambda \quad (8)$$

For example, if a band had a transmission of 0.7 over a span of 10 nm, it would have an equivalent width of 3 nm. We convert our filter-averaged transmissions to $W_{gas}$ to facilitate comparison with prior works, which were often reported as Equivalent Widths. When the transmissions are close to unity, as they are here for the 619 nm and 647 nm molecular bands, and the opacities are thus small, the Equivalent Width of the absorption band may be estimated from the filter-averaged opacity, $\bar{\tau}$, by the relation:

$$W_{gas} \approx \bar{\tau} \times \frac{\int_{\lambda_1}^{\lambda_2} a_{gas}(\lambda) d\lambda}{k_{eff}} \approx (1 - \bar{T}) \times \frac{\int_{\lambda_1}^{\lambda_2} a_{gas}(\lambda) d\lambda}{k_{eff}} = (1 - \bar{T}) \times C \quad (9)$$

where $a_{gas}(\lambda)$ is the gas absorption coefficient spectrum from Fig. 1c and $C$ is the proportionality constant represented by the integral divided by $k_{eff}$ (see the Appendix for details). Table 4 shows the transmission, opacity, and equivalent widths for the methane and ammonia bands averaged over the disk of Jupiter.

**Table 4:** Disk average methane and ammonia absorption properties as determined by convolving SCT filter profiles with observed VLT/MUSE albedo spectra and the Karkoschka [1994] albedo spectrum (Figure 1).

| Parameter | MUSE 2022-07-30 | MUSE 2022-09-19 | MUSE Avg. | Karkoschka (1994) |
|---|---|---|---|---|
| $\bar{T}_{619}$ | 0.896 | 0.899 | 0.897 | 0.884 |
| $\bar{\tau}_{619}$ | 0.110 | 0.107 | 0.108 | 0.124 |
| $W_{619}$ (nm) | 1.336 | 1.296 | 1.316 | 1.494 |
| $\bar{T}_{647}$ | 0.960 | 0.969 | 0.964 | 0.963 |
| $\bar{\tau}_{647}$ | 0.041 | 0.031 | 0.036 | 0.037 |
| $W_{647}$ (nm) | 0.486 | 0.371 | 0.429 | 0.440 |

By comparing the two MUSE observations and the Karkoschka result, we gain insight into the observational variations and uncertainties. For transmission, the variation, $\delta T/T$ is about 0.75% for CH$_4$ and 0.40% for NH$_3$. Propagating the the transmission variation into opacity we have:

$$\frac{\delta\tau}{\tau} = \frac{-\ln(T \pm \delta T) + \ln(T)}{\ln(T)} \qquad (10)$$

Numerically, this results in $\delta\tau/\tau \approx \pm 6\%$ error for methane and $\delta\tau/\tau \approx \pm 11\%$ for ammonia. This range of variation applies to equivalent width since it is linearly related to opacity.

Many spectroscopic measurements of the ammonia equivalent width at 647 nm have been made over past decades [*Atai et al.*, 2022; *Moreno and Molina*, 1991; *Moreno et al.*, 1988; *Sato and Hansen*, 1979; *Teifel' et al.*, 2018; *Vdovichenko et al.*, 2021]. The extremes range from 0.3 nm to 1.29 nm for different locations on the disk at different epochs. A similarly extensive set of measurements exists for the 619 nm band of methane falling in a range of 0.91 to 2.25 nm [*Lutz and Owen*, 1980; *Lutz et al.*, 1982; *Moreno and Molina*, 1991; *Moreno et al.*, 1988] with most falling close to 1.6 nm. Our models and observations fall well within the envelope of prior analyses.

With the opacities for methane and ammonia, we can now compute $N$ from disk-averaged observations (for which $\eta$=4) and then determine $P_{cloud}$ (computed from $N_{CH4}$) and $\bar{f}_c(NH_3)$ (Table 5). The opacity uncertainty estimates described carry forward into these parameters.

**Table 5:** Disk-Integrated environmental parameters: column abundance, effective cloud-top pressure, and ammonia mole fraction.

| Parameter | MUSE 2022-07-30 | MUSE 2022-09-19 | MUSE Avg. | Kark |
|---|---|---|---|---|
| $N_{CH4}$ (m-amagat) | 258 | 250 | 254 | 290 |
| $P_{cloud}$ ($\eta$=4) (mb) | 803 | 777 | 790 | 903 |
| $N_{NH3}$ (m-amagat) | 14.0 | 10.6 | 12.3 | 12.6 |
| $f_c$(NH$_3$) (ppm) | 111 | 87 | 99 | 89 |

The idea of "cloud tops" is a convenient simplification to describe aerosol scattering in Jupiter's atmosphere which uses a simple single reflecting layer model. In current models, multiple cloud and haze layers of varying opacity and vertical extent along with coloring agents (chromophores) are considered [*Wong et al.*, 2023]. However, the scattering path above the 'cloud tops' should be roughly commensurate with cloud layers in more sophisticated models. A review of cloud models [*Dahl et al.*, 2021] show high opacity 'sheet cloud' pressures ranging from 2.0 bar to 670 mb, for different Jovian features. For vertically extended clouds, base pressures extend from nearly 4.9 bars to 2.2 bars and cloud tops from 490 to 60 mb, again for different features. In all cases overlying haze and coloring layers are assumed. In the simplified RLM analysis performed here, the range of results from ~870 mb to ~1.0 bar is in reasonable agreement with the literature.

The ammonia total column abundance, $N$, can also be compared to the literature. Using the 647 nm band, NH$_3$ values of 12±1.5 m-amagat were found by Encrenaz et al. [1974]. Woodman et al. [1977] find 13.1±0.7 m-amagat for NH$_3$ and 195±10 m-amagat for CH$_4$ using the 647 and 619 nm bands. This C/N ratio gives an ammonia column abundance range of 110-135 ppm. Using the 542 and 552 nm bands, they find larger values of 200-280 ppm, however they attribute this to the fact that this band-pair is weaker than the 619-647 nm pair and thus penetrates deeper into the atmosphere where C/N ratio is smaller. An ammonia mole fraction of about 170 ppm can be found from C/N~10.8 [*Lutz and Owen*, 1980] and Kunde et al. [1982] find a value of ~180 ppm sharply decreasing above 800 mb. Moeckel et al. [2022] show values well below 100 ppm at altitudes above about 500 mb. Given the variety of observations and retrieval methods, our result of ~90 ppm for the column-averaged mole fraction seems within reason.

SCT transmissions for methane and ammonia are calibrated by normalizing to the transmission of the MUSE average values shown in column 4 of Table 4. While there may be real variations in the disk average transmission over time, the relative spatial variations in terms of meridional profiles and two-dimensional maps will still be well represented, but may have an offset in absolute calibration.

## 5 Meridional Profiles

The logical next step is to look at meridional profiles of ammonia absorption and abundance. There are long-running records of absorption profiles along with a number of abundance measurements determined at specific pressure levels.

### 5.1 Absorption

Absorption profile data is readily available in the literature, often in the form of equivalent width of the ammonia absorption band as a function of latitude [*Atai et al.*, 2022; *Teifel' et al.*, 2018; *Vdovichenko et al.*, 2021]. In this work we compute the average transmission at each latitude to create meridional profiles. We then translate our transmission into equivalent widths (Section 4) for comparison to prior works.

Figure 3 compares ammonia absorption profiles generated from SCT and VLT/MUSE data from 2020 to 2023 to spectral observations made in 2016 with a 0.6 m telescope [*Teifel' et al.*, 2018]. The SCT and VLT/MUSE observations were averaged along lines of latitude within ±1º of longitude from the central meridian. The spectral observations were made with a 25-µm slit [inferred from the instrument configuration described by *Teifel' et al.*, 2018] which covers ~2º of longitude on Jupiter's disk when aligned along the central meridian.

Large scale similarities in the profiles' morphologies include the northern EZ enhancement and NEB-NTrZ depletion in absorption. Also, there is a broad, less intense enhancement from the southern EZ through the SEB seen in the 2022-2023 SCT and MUSE data as well as the 2016 *Teifel et al.* [2018]. However, this enhancement is not seen in the 2020-2021 SCT data and may indicate change over time. Smaller scale features are evident, particularly in the VLT/MUSE profile for 2022 and some are echoed, albeit less sharply, in the SCT 2022 data. These features include enhanced absorption in the STZ and STB regions, the SEB, and the NTZ.

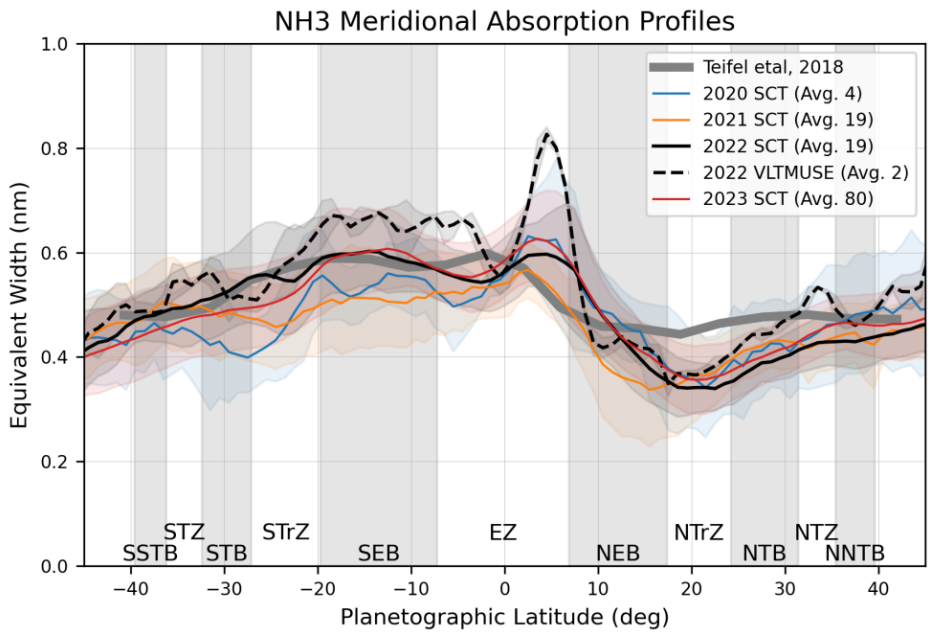

**Figure 3:** Meridional profiles of NH$_3$ absorption (647 nm) equivalent width (nm) comparison: SCT observations from the 2020-2023 apparitions per Table 1, the average of the two VLT/MUSE observations described in the text, and a 2016 profile (average of 12 obs.) from long slit spectra [*Teifel et al.*, 2018].

The observations from 2020-2022 consistently show the sharp northern EZ enhancement, which is absent from the 2016 observations. This may represent a real physical change related to the ochre coloration of the EZ extending from 2019 through early 2022. Year-to-year changes may exist in the 2020-22 data, but the signal-to-noise ratio (SNR) is relatively low so potential changes are difficult to identify with any confidence. One possible cause is that the fading of the visually dark NEB from 2020-21 [*Rogers et al.*, 2022] has influenced the shape of the absorption curves in that region. During 2015-16 the dark NEB was undergoing an expansion event [*Fletcher et al.*, 2017; *Rogers et al.*, 2019] and that curve [*Teifel' et al.*, 2018] shows the flattest absorption in the NEB and NTrZ. However, with only ammonia absorption data, it is not possible to distinguish between cloud height changes and ammonia abundance changes.

5.2 Abundance

Meridional profiles of ammonia absorption represent a complex mix of spatial variations of ammonia, aerosols in the form of clouds, hazes, and chromophores, Rayleigh scattering, and illumination and viewing geometry. Comparing to the known methane abundance along a near-identical scattering path allows retrieval of the spatial variations of ammonia abundance (Eq. 7) while (nearly) eliminating all the other effects. Figure 4 shows the results of this work for the 2022 VLT/MUSE and 2022-2023 SCT column-averaged abundances, compared to March 2017 NH$_3$ abundances at 574, 657, and 752 mb, determined from Gemini/TEXES mid-infrared spectrometer observations [*Fletcher et al.*, 2020]. The SCT and VLT/MUSE data represent longitudinal averages within ±1° of the central meridian. The SCT data represents the average of seven profiles, and the envelope (standard error) of their variation is shown. Like the absorption profile, the SCT mole fraction curve echoes the VLT/MUSE curve, but shows less contrast. This may be due to the better spatial resolution of the VLT/MUSE data.

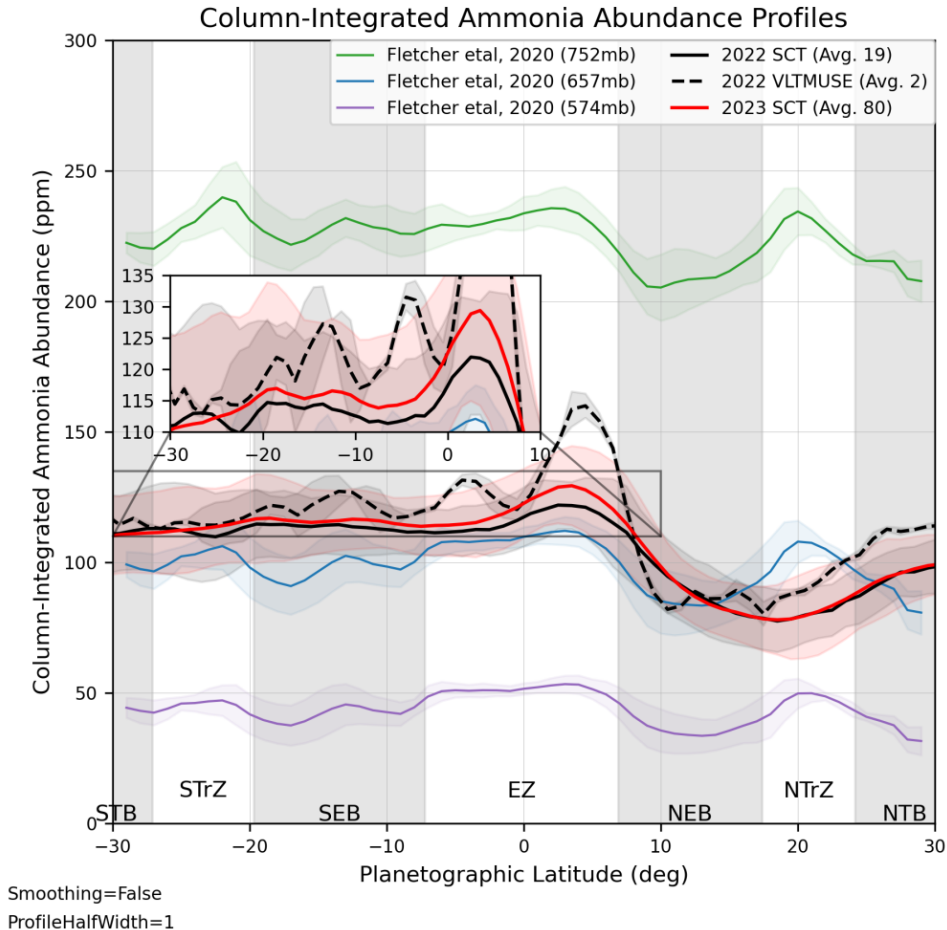

**Figure 4:** Meridional profiles of ammonia mole fraction: 2022-2023 SCT column-averaged mole fractions, VLT/MUSE 2022 column-averaged mole fractions; 2017 March 12–14 Gemini/TEXES data retrieval at three pressure levels (average and variation of seven observations from supplementary data of *Fletcher et al.*, 2020). The SCT and VLT/MUSE profiles are 2° wide along the central meridian (see text for details).

For the mid-IR Gemini/TEXES observations [*Fletcher et al.,* 2020], supplementary data provided meridional profiles retrieved using NEMESIS [*Irwin et al.,* 2008] for each of seven observing groups at 61 different pressure levels. The seven-group average and variation (standard error) at the 657 mb pressure level most closely matches the value of column-averaged mole fraction from the current work. Note that the 657 mb data is shown primarily for illustrative purposes, since it represents the mole fraction at a specific pressure level, and the SCT and MUSE data represents average mole fraction *above* the effective cloud-top pressure level.

Given the different nature of the NEMESIS retrieval, we cautiously note similarities with the SCT and VLT/MUSE curves including the enhanced $NH_3$ in the northern EZ and the depletion seen in the NEB. Additionally, there is notable structure is seen in both 2017 and 2022-2023 in the SEB (Fig. 4 inset).

Differences are also evident. First, in 2022-2023, the depletion in the NEB extends through the NTrZ whereas in 2017 an enhancement is seen in the NTrZ. The difference could be related to major atmospheric changes observed in the visible waveband: in March 2017 the dark NEB was broad and the NTrZ was also darkened following a vigorous outbreak on the NTBs jet [*Rogers et al.*, 2019]; whereas in 2022 the NTrZ and the northern part of the NEB were quiet and largely white, consistent with thicker cloud cover [*Rogers et al.*, 2022].

The southern half of the EZ is relatively depleted in 2022-2023, and in 2017 the EZ enhancement is broad and flat, with only a minor relative peak in the northern EZ. Finally, the 2022 data show the STrZ to be flat or depleted while the 2017 data show a peak in this zone.

## 6 Ammonia in Two Dimensions

The observations obtained with the SCT along with those from VLT/MUSE were of sufficient quality to produce two-dimensional maps of the column-averaged ammonia abundance. The process outlined in Section 3 was followed the same way as in the disk-integrated and the meridional analyses above. As noted earlier in this paper, ammonia abundance relative to methane abundance decreases with altitude in addition to being spatially variable. It is necessary then to understand the degree to which the column-averaged ammonia abundance is a secondary effect of cloud-top pressure versus being dominated by its horizontal variations.

We now consider the relationship between cloud-top pressure and ammonia abundance above the clouds. Figure 5 presents the ammonia abundance and effective cloud-top pressure for 2022-10-09 04:01:30, one of the sharpest observations made with the SCT.

The lowest pressure cloud-tops are clearly associated with the NTrZ and the whitened NTB, where there is a gradient of ammonia abundance. A general increase in cloud-top pressure is seen moving southward, with local maxima in the NEB and local minima in the EZ. It may appear surprising that the NEB does not exhibit even lower cloud heights, but this probably reflects its unusual state in 2022 [*Rogers et al.*, 2022]. There is a local minimum of ammonia in the NEB and a local maximum in the northern EZ. The highest cloud-top pressures are seen in the SEB where the ammonia abundance is at a relative minimum.

Beyond the variations with latitude, one can see localized ammonia enhancements associated with several small bright regions in Fig. 5 (arrows 1-4). These regions do not show consistent differences in cloud-top pressure, but in some cases like arrowed region 1 there is a small decrease in pressure. This suggests that these bright areas are associated with ammonia transport or convergence either vertically or horizontally. They are consistent with the $NH_3$ enhancements in the northern EZ, sometimes referred to as plumes, that have been reported from observations in the thermal infrared [*Fletcher et al.*, 2020] and radio [*de Pater et al.*, 2016; *de Pater et al.*, 2019] in localized patches, and from the Juno Microwave Radiometer circumglobally [*Bolton et al.*, 2017; *Bolton et al.*, 2021].

The ammonia enhancements also overlap some of the North Equatorial Dark Features (NEDFs) adjacent to plumes. But it has been noted that plumes are not uniformly enhanced in $NH_3$ nor are NEDFs uniformly depleted in $NH_3$. "In one extreme case (DF6), the whole eastern edge of a DF was hidden by a region of enriched NH3 gas, stretching from the northern EZ into the NEB" [*Fletcher et al.*, 2020]. Additionally, Bjoraker *et al.* (2022) show that 5 μm hot spots do not exhibit minima (or maxima) in $NH_3$. Finally, regions of high $NH_3$ have also been detected within NEDFs (or 5 μm) in infrared spectral data from Galileo and Juno [*Roos-Serote et al.*, 1998; *Grassi et al.,* 2017].

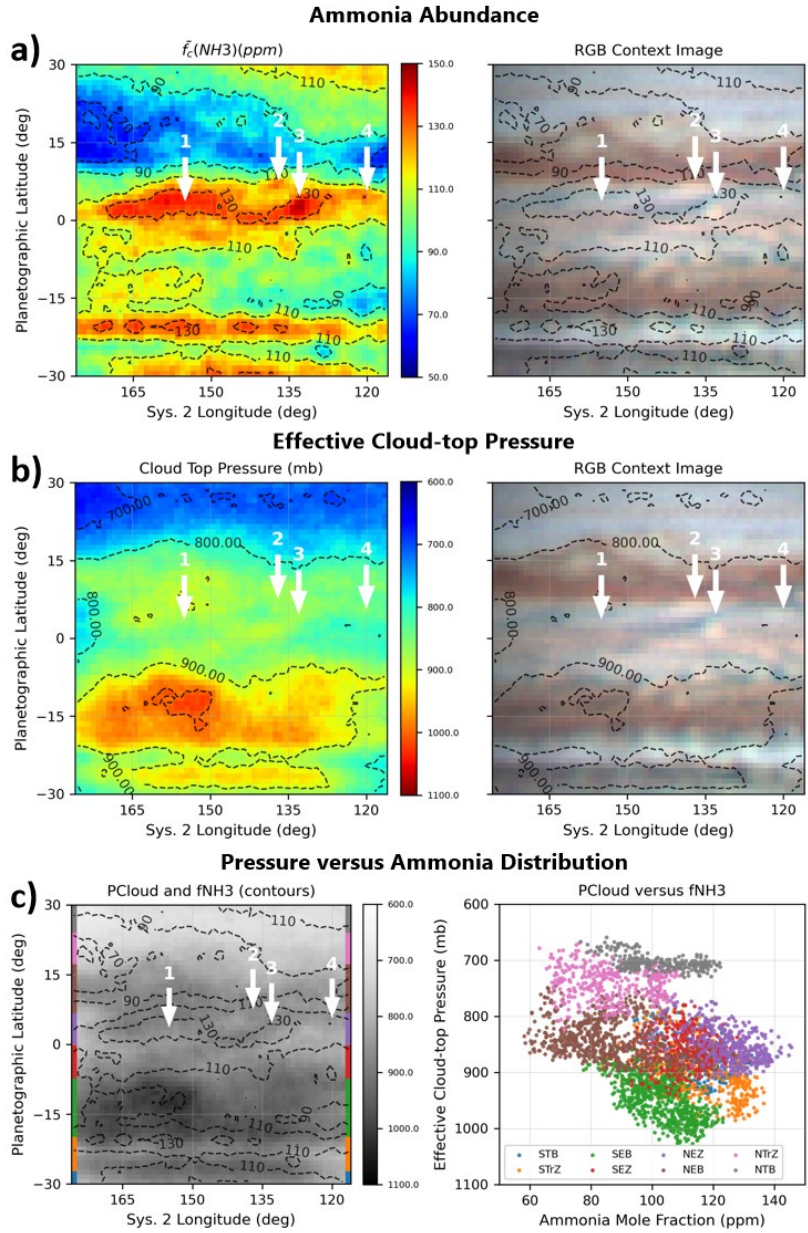

**Figure 5:** Ammonia mole fraction compared to cloud-top pressure estimated from an SCT observation at 2022-10-09 04:01:30Z. a) Column-averaged ammonia mole fraction. Arrows show selected ammonia enhancements in the northern EZ. b) Estimated cloud-top pressure using methane absorption and a reflecting layer model assumption. c) Contours showing $\bar{f}_c(NH_3)$ on top of a grey-scale representation of cloud top pressure and a scatter plot of column-averaged ammonia mole fraction above the cloud-tops versus cloud-top pressure Individual zonal bands are identified in different colors. See text for discussion of the scatter plot clustering.

We further evaluate the relationship between ammonia abundance and cloud-top pressure in Fig. 5c. First, we plot contours of $\bar{f}_c(NH_3)$ over a greyscale representation of $P_{cloud}$ over the region depicted in Fig. 5. Then we look for correlations by creating a scatter plot of the data.

Specifically, we plot the value of ammonia mole fraction versus the cloud-top pressure for every 1 x 1 degree element in Fig. 5. In addition, the points are color-coded by band. This plot shows a weak positive correlation pressure and ammonia mole fraction, but that correlation is dominated by the associations with band/feature type. However, unique clustering of the data is seen for each of the meridional bands. This clustering is evidence that the ammonia abundance is indeed dependent on feature types that are mostly associated with latitude, but not necessarily associated with cloud-top pressure. Further investigation is warranted to determine what attributes of these bands drives the ammonia variation if not cloud height.

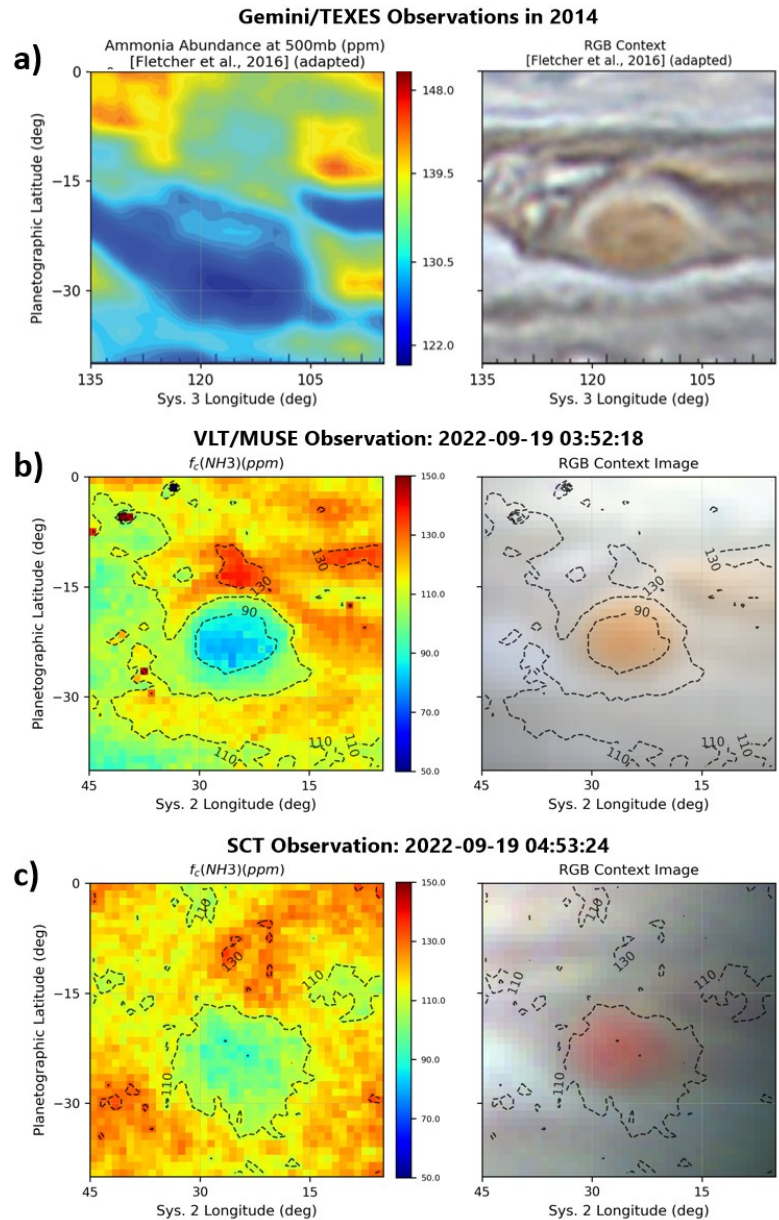

**Figure 6:** Ammonia abundance and visual appearance: a) Gemini/TEXES determined abundance in December 2014 at 500mb, adapted (cropped to the same latitude and longitude extent as VLT/MUSE and SCT panels) from Fletcher et al. (2016); b) VLT/MUSE-determined column-average abundance: 2022-09-19 0352UT (Sys. II CML 13.5); c) SCT column-average abundance: 2022-09-19 0453UT (Sys. II CML 50.4).

Having shown that there must be aspects of ammonia variation not correlated with cloud height, we now examine the GRS and its surrounding region. The GRS is chosen as a mostly stable feature expected to show ammonia depletion. Figure 6 shows three pairs of images. On the left are maps depicting ammonia mole fraction (retrieved at a specific pressure level or column-averaged) and on the right are maps made from contemporaneous visual images for orientation.

Figure 6a is adapted from a NEMESIS retrieval of $NH_3$ abundance at 500 mb using IRTF/TEXES mid-infrared observations made in December 2014 [*Fletcher et al.*, 2016]. The most notable feature is the depletion associated with the GRS. The apparent offset of this depletion to the south of the GRS visual center by about 6 degrees was noted, but no physical mechanism could be identified as an explanation [*Fletcher et al.*, 2016].

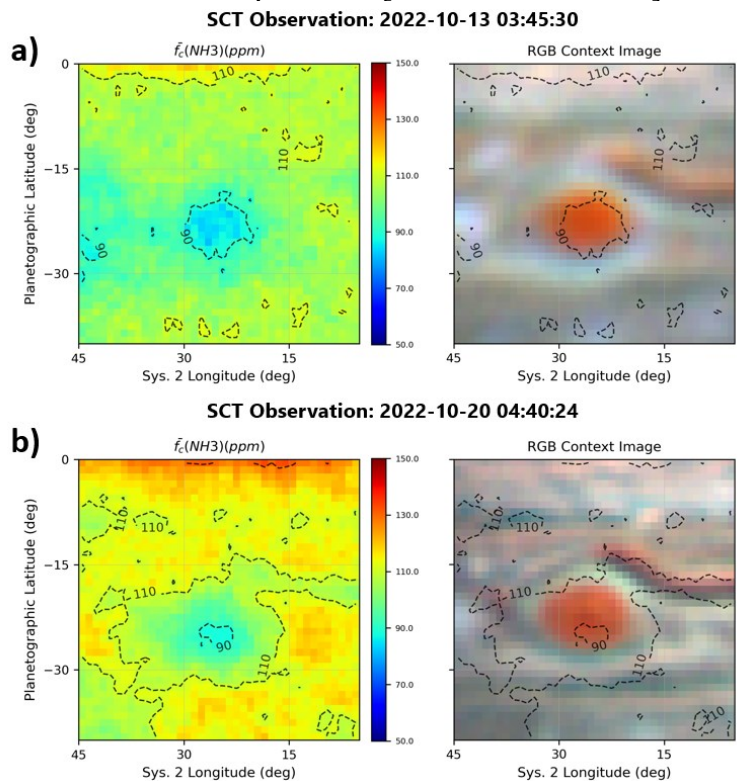

**Figure 7:** SCT-determined column-averaged ammonia mole fraction in the GRS region separated by one week in time. Top: 2022-10-13 03:45UT (Sys. II CML 18.7); bottom: 2022-10-20 04:40UT (Sys. II CML 24.3). Large scale features persist, but smaller scale features show changes. While these changes may be physical, they may also represent indications of noise.

Figure 6b shows a map of column-averaged ammonia abundance determined from the VLT/MUSE data on 2022-09-19. Broad similarities are seen in the depletion near the red spot, including its bias to the south by about 2º. In addition to the GRS depletion, there are notable $NH_3$ variations across the surrounding belts and zones, substantially reproducible within the 2022 observations, but somewhat different in the 2014 Gemini/TEXES retrieval probing the 500 mb level. Figure 6c demonstrates that the SCT observations from a single night are sufficient to show large-scale, strong features like the GRS depletion, including the southward offset. Smaller or weaker features that are seen will be more impacted by statistical noise, though smoothing or filtering could reduce that impact. Continuing the examination of the GRS region, two higher quality observations were obtained one week apart in October 2022 (Fig. 7). The GRS depletion and a portion of the EZ enhancement were very clear in both observations. Of interest is the persistence of almost all lesser features along with their evolution during the week. As single observations, noise must be considered when interpreting the images.

    Finally, we look at apparition averages of the GRS region in 2022 and 2023 (Fig. 8). The GRS depletion is clear, though somewhat less intense in 2023 than in 2022. Likewise the EZ enhancement is very apparent. Ammonia is slightly enhanced in the STrZ both years. The SEBs depletion appears more distinct in 2023 than in 2022. Patchy enhancement (arrow 2) is seen north of the Red Spot Hollow (RSH, arrow 1) both years. The features seen are substantially the same as those in the MUSE data shown in Fig. 6b, though they vary in detail. A portion of the differences seen between 2022 and 2023 may be attributable to differences in the number of observations averaged (five versus ten) and the period of time averaged over (two months versus four months)

    Notably, in both years the center of the GRS ammonia depletion is offset southward from the visual center of the GRS by about 2º. This is consistent with the VLT/MUSE observations and the single-image SCT observations in 2022. This southward offset is also qualitatively consistent with the retrieval by [*Fletcher et al.*, 2016] (our 7a), and the Juno Microwave Radiometer data above the ~1.5-bar level [*Bolton et al.*, 2021]. While Bolton et al. (2021) highlighted $NH_3$ enhancement in the northern half of the GRS, they noted its spatial heterogeneity and their Figure S4 shows that there is strong depletion in its southern half and southern collar.

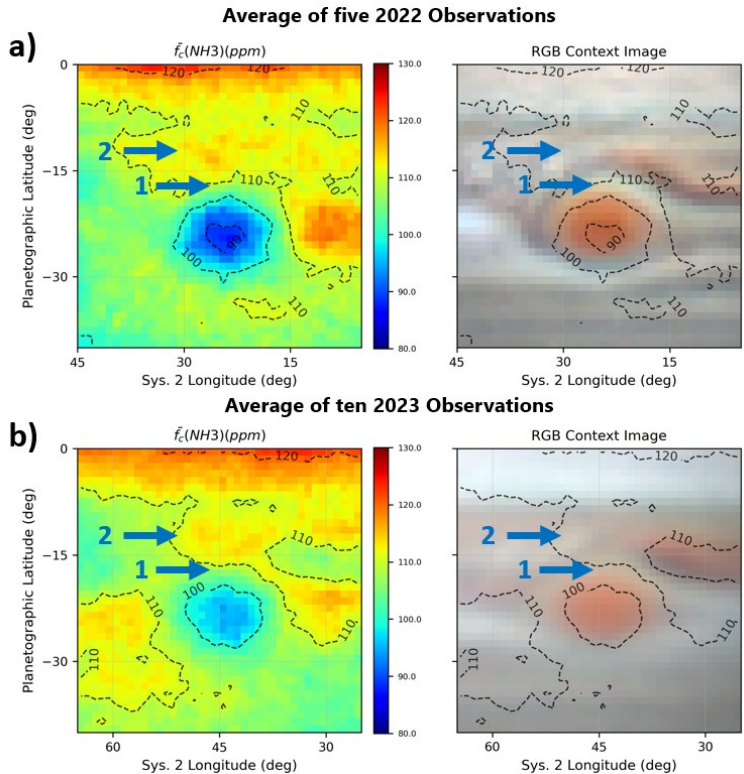

**Figure 8:** Apparition averages of the GRS region ammonia abundance in 2022 (Aug. 18, 28, Sep. 04, Oct. 13, 20) and 2023 (Aug. 31 – two observations, Sep. 5, 24, Oct 5, 17, 22, Nov. 13, Dec. 7). On these dates, the GRS was observed within 17.7 º of the central meridian (except in one case where the offset was 32.1 º) and were checked for quality. The scale has been stretched to highlight more subtle features compared to the other ammonia maps in this work. Arrowed features are described in the text.

## 7 Summary and Conclusion

This work has adapted a spectroscopic approach outlined in the 1970s for determining ammonia absorption in Jupiter's upper troposphere from small-telescope, narrow band imaging. Using spectrally adjacent bands of ammonia and methane, we minimize differential effects of Rayleigh and aerosol scattering and spectral reflectivity. Since methane abundance is fixed, it can be used as a benchmark against which ammonia can be measured.

In the process, this work has compared ammonia absorption in disk-integrated and meridional profiles to prior work. Then ammonia abundances are determined for meridional profiles and resolved maps of Jupiter. The results are compared to ammonia abundance retrievals from the TEXES MIR spectrometer observations at the Gemini and IRTF telescopes. In addition, using a VLT/MUSE data cube we created images for analysis, which compare favorably to the SCT data.

Features seen in the column-averaged ammonia abundance include the EZ enhancement, NEB depletion, the depletion in the GRS, and smaller-scale enhancements and depletions apparently associated with cloud structures. Notably, we provide additional evidence for a southward offset of the GRS ammonia depletion. Interpretation of our results, especially in the context of full atmospheric retrievals, is beyond the present scope, though we note that significant visual changes in the atmosphere occurred between our observations and the comparison data from 2014-2016. The accumulation of additional observations will support investigation of short term (days to weeks) variations in ammonia as they relate to changes in Jupiter's large-scale visible cloud structures that would otherwise be difficult to achieve with only professional ground-based observations. Observations will be compared to professional observations periodically to validate the calibrate amateur data.

With the ability to obtain a meaningful measure of ammonia (column-averaged mole fraction), we now seek more frequent routine monitoring by encouraging and enabling advanced amateurs to monitoring for short-term changes and respond to meteorological events.  Example phenomena of interest include the North Equatorial Dark Features that can evolve rapidly, the GRS as it encounters other features or exhibits visible changes, jetstream outbreaks in the belts, and NEB expansion events. The filters needed are economical in the sense that they cost only a fraction of the base system needed to conduct the observations (~0.25 m telescope and CMOS camera). In addition, we seek to obtain higher resolution and SNR by accessing instruments up to 0.5-m class as well as data sharing with large programs like VLT/MUSE observations where feasible.

**Open Research**

Datasets for this research, including figures, FITS files, and spectral data files are available from Hill (2024). Data for ammonia abundance profile comparisons used in Fig. 4 from Fletcher et al. (2020) are available from Fletcher (2020).

**Acknowledgments**

**Appendix**

We seek to compare bandpass filter measurements of molecular band opacity with historical measurements that are represented as equivalent width:

$$W = \int 1 - T(\lambda) d\lambda$$

where $T$ is transmission. To do this we establish an approximate linear relationship between opacity, $\tau$, and equivalent width. Two assumptions are necessary:

1. Opacity, $\tau(\lambda)$, is small, and
2. Absorption cross section, $a_{gas}(\lambda)$, is constant, e.g., not dependent on pressure or temperature.

Since transmission and opacity are related as $T(\lambda) = e^{\tau(\lambda)}$, if opacity is small then $\tau(\lambda) \approx 1 - T(\lambda)$ and we can write:

$$W \approx \int \tau(\lambda) d\lambda$$

The filter-averaged opacity is:

$$\bar{\tau} \approx \frac{\int f(\lambda)\tau(\lambda) d\lambda}{\int f(\lambda) d\lambda}$$

where $f(\lambda)$ is the bandpass filter profile. We know that the opacity at a given wavelength is dependent on the product of $a_{gas}$, the wavelength dependent absorption cross section, $N_{gas}$, the column number density of the gas, and $\eta$, the airmass factor: $\tau(\lambda) = (\eta N_{gas}) a_{gas}(\lambda)$. Thus, we can express the filter-averaged opacity:

$$\bar{\tau} = \frac{(\eta N_{gas}) \int f(\lambda) a_{gas}(\lambda) d\lambda}{\int f(\lambda) d\lambda}$$

We can also compute the filter-averaged absorption coefficient, $k_{eff}$, as:

$$k_{eff} = \frac{\int f(\lambda) a_{gas}(\lambda) d\lambda}{\int f(\lambda) d\lambda}$$

then filter-averaged opacity becomes: $\bar{\tau} = (\eta N_{gas}) k_{eff}$. Since our goal is to find a relationship between filter-averaged opacity and the molecular band equivalent width, we examine:

$$\frac{\bar{\tau}}{W} \approx \frac{\int f(\lambda)\tau(\lambda) d\lambda}{\int f(\lambda) d\lambda} \cdot \frac{1}{\int \tau(\lambda) d\lambda} = \frac{(\eta N_{gas}) \int f(\lambda) a_{gas}(\lambda) d\lambda}{\int f(\lambda) d\lambda} \cdot \frac{1}{(\eta N_{gas}) \int a_{gas}(\lambda) d\lambda}$$

which can be reduced to the following relationship and shown to be approximately constant:

$$\frac{\bar{\tau}}{W} \approx \frac{k_{eff}}{\int a_{gas}(\lambda) d\lambda} = C$$

because $k_{eff}$ and $\int a_{gas}(\lambda) d\lambda$ are not dependent on the amount of absorbing gas given the starting assumption of small opacity. While the value of the proportionality constant C may change depending on the filter profile, the relationship between opacity and equivalent width is approximately linear for small values of opacity.

$$\bar{\tau} \approx C \cdot W$$

We compute the proportionality constant, $C$, for a grid of abundances from 0-500 m-amagat for methane and 0-30 m-amagat for ammonia, which fully encompasses the range of column densities expected above Jupiter's clouds. Over these ranges, the proportionality constants for methane and ammonia are 12.86 nm$^{-1}$ and 12.05 nm$^{-1}$.